\documentstyle[12pt,aaspp4]{article}
\begin{document}

\title{Anisotropies in the Motions and Positions of the Galactic Globular 
Clusters}

\author{F.D.A. Hartwick}

\affil{Department of Physics and Astronomy, \linebreak University of Victoria, 
Victoria, BC, Canada, V8W 3P6}

\begin {abstract}
The velocity ellipsoid for 38 globular clusters with [Fe/H]$\leq-1.0$ is
derived and shown to be significantly anisotropic with major axis directed
towards low Galactic latitude.  Principal axes of the spatial distribution
of different groups of clusters are derived and compared with the velocity
ellipsoid.  The metal poor cluster spatial distribution is significantly 
flattened along an axis which coincides within the uncertainties with the 
major axis of the velocity ellipsoid.  Given the observed steep age-metallicity
relation for metal poor clusters, one speculative interpretation of the
data is that an initially flattened filament underwent a relatively rapid
initial transverse collapse forming satellite galaxies and metal poor
globular clusters while the protogalaxy collapsed and assembled more
slowly along the filament acquiring and/or redistributing angular momentum in 
the process. 
\end{abstract}


\section {Introduction}

Globular clusters provide a unique probe of the earliest phases of the
evolution of the Galaxy.  From the early work of Kinman (1959),
Zinn (1985), and others it is clear that the clusters can be divided into at
least two groups -- the spatially extended, metal poor blue clusters and
the more centrally concentrated towards the Galactic center, metal rich red 
clusters.  This work is
concerned primarily with the former group, and the main motivation is the
recent availability of space motions for a significant number of these
clusters as published by Dinescu et al.\ (1999). Dinescu (private 
communication) has also provided
this author with data for 4 additional clusters. These data are used to 
determine the velocity ellipsoid for the metal poor group in order to allow 
a comparison  with the spatial distribution. In a previous paper (Hartwick 
2000, hereafter H2000) the spatial distribution of various halo samples was 
examined. There it was found that the metal poor group of globular clusters
formed an oblate distribution whose minor axis was directed to a low Galactic 
latitude. This was in contrast to the metal rich group which exhibited a 
triaxial distribution with minor axis toward the Galactic pole. We re-examine  
the spatial distributions with modified samples of clusters and compare the 
results with the kinematics.

\section{The Velocity Dispersion Tensor of the Halo Clusters}

Danescu et al.\ (1999) have compiled space motions for 38 globular
clusters, and 35 of these have [Fe/H] values less than $-1.0$.
Since that time 3 more halo clusters (NGC6522, NGC7006, \& Pal 13) 
have been added to the compilation.
This sample of 38 clusters is the basis for the following kinematical 
investigation.  The tabulated
UVW components of space motion (corrected solar peculiar motion and for
Galactic rotation of 220 km~sec$^{-1}$) are referred to to the Galactic
center where $V_{\pi}$ is the radial component in the Galactic plane
directed positive away from the center, $V_{\phi}$ is the tangential
component positive in the direction of Galactic rotation and $V_{z}$ is
the vertical component directed positive in the direction of the north
Galactic pole. 

The average of each of these components for the entire sample was removed
before forming the following six moments: $\overline{V_{\pi}V_{\pi}},
\overline{V_{\phi}V_{\phi}}, \overline{V_{z}V_{z}},
\overline{V_{\pi}V_{\phi}}, \overline{V_{\pi}V_{z}}$, and
$\overline{V_{\phi}V_{z}}$.  These six moments are the components of the
symmetrical velocity dispersion tensor.  The eigenvalues of the matrix
were then determined in the standard way and are given below.  The units
of the eigenvalues denoted by $\sigma$ are km~sec$^{-1}$ and the one sigma
uncertainties throughout were computed by the bootstrap method. 

\begin{center}
Clusters with [Fe/H]$<-1.00$\ (n=38)
\end{center}
\vspace{-0.3 in}
\begin{eqnarray}
\sigma_{a} = 161^{+18.3}_{-15.0} & l = 338^{+10.5}_{-6.65} & b = 
-5.65^{+13.6}_{-15.0} \nonumber \\
\sigma_{b} = 120^{+9.44}_{-15.5} & l = 87.2^{+155.}_{-30.0.} & b = 
-73.5^{+27.9}_{-6.63} \\
\sigma_{c} = 95.6^{+2.66}_{-17.6} & l = 66.1^{+11.2}_{-10.8} & b = 
15.4^{+22.2}_{-21.1} \nonumber \\
& \overline{V_{\phi}} = 59.6^{+13.5}_{-17.9} \nonumber
\end{eqnarray}

\noindent
As a check on the sensitivity of the above result to the [Fe/H] cutoff, we
repeated the procedure for the 26 clusters with [Fe/H]$<-1.5$ and find: 

\begin{center}
Clusters with [Fe/H]$<-1.50$\ (n=26)
\end{center}
\vspace{-0.3 in}
\begin{eqnarray}
\sigma_{a} = 172^{+17.9}_{-19.2} & l = 330^{+11.4}_{-8.11} & b = 
-6.31^{+14.9}_{-14.5} \nonumber \\
\sigma_{b} = 122^{+8.45}_{-27.7} & l = 179^{+91.7}_{-112.} & b = 
-82.8^{+32.6}_{+6.58} \\
\sigma_{c} = 94.7^{+5.76}_{-27.8} & l = 60.8^{+12.1}_{-12.1} & b = 
-3.48^{+30.1}_{-21.3} \nonumber \\
& \overline{V_{\phi}} = 49.3^{+23.7}_{-28.1} \nonumber
\end{eqnarray}

In both of the above solutions the mean streaming motions in the $\pi$ and $z$
directions differed from zero by less than 1.5 sigma. Also the above values of 
$\overline{V_{\phi}}$ are independent of distance from the Galactic center.

Asymmetric error ellipsoids in the components of space motion can be expected
due to relatively small errors in radial velocity and relatively large errors
in proper motion. In order to ensure that such effects were not unduly 
influencing solution (1) the errors in radial velocity, proper motion and 
distance given by Dinescu et al.\ (1999) were used as dispersions in assumed 
gaussian distributions and solutions were performed on space motions 
determined by the propagation of randomly chosen uncertainties in each of the 
above quantities. The mean and standard deviation of the major axis from 100 
trials with 38 clusters is $\sigma_{a} = 167 \pm 7.98$ towards $l = 332 \pm 
6.28$ and $b = -5.94 \pm 7.64$. Results for the other two axes are $\sigma_
{b} = 122 \pm 10.9$ and $\sigma_{c} = 108 \pm 10.1$. This independent 
assessment of uncertainties thus strengthens the case for the generally 
$prolate$ velocity ellipsoid given by solution (1).

An examination of the spatial distribution of the clusters from solution 
(1) shows that only 7 of the 38 clusters were in the hemisphere
beyond the Galactic center. In order to assess the effect of this spatial 
imbalance on our result, kinematic
solutions were made using the above 7 clusters combined with 7 independently
chosen clusters from the hemisphere containing the sun. Thirty solutions 
were made with 14 clusters each and the median results for each aspect of 
the major axis are: $\sigma_{a} = 166^{+20}_{-9},\ \ l = 348^{+8}_{-7},\ \ $ 
and $b = -13^{+29}_{-8}$. 
While the one sigma uncertainties of solution (1) do overlap with the above 
result, perhaps the most important implication of this exercise is a clear 
need for more cluster space motion determinations.

\section{The Spatial Distribution of the Globular Clusters}

The spatial distribution of the outlying satellites and the globular
clusters was determined in H2000.  There it was found that the 99 metal
poor globular clusters ([Fe/H]$\leq-1.0$) formed an oblate distribution
with minor axis directed towards $l=307^{\circ}$,$b=-3.3^{\circ}$ quite
different from the short axis of the distribution of the satellite
galaxies.  The result from the kinematics described above prompted a
re-examination of this result. It is believed that the pole of the orbit
of the Sagittarius system is highly inclined to that of the remaining
outlying satellite galaxies (Lynden-Bell \& Lynden-Bell 1995).  As well
these authors tentatively assign the 6 clusters M54, NGC2419, Arp 2, Pal 2, 
Terzan 7 and Terzan 8 to the Sagittarius stream.  Given that the present 
orbit of
Sagittarius is probably not the one that it had at formation (c.f. Zhao
1998, also Johnston et al. 2002) we removed the above clusters from the
original sample and performed new solutions. (Note that the Harris (1996) 
web-based compilation contains 147 clusters of which 99 have [Fe/H]$\leq-
1.0$. Culling 5 clusters from this sub-group still leaves us with 95\% of the 
original sample. Furthermore, none of these five clusters was included in 
the above kinematic solutions.) The analysis procedure followed here is
slightly different from that in H2000.  Here the tensor components
$\overline{x_ix_i}, \overline{x_iy_i}$ etc were calculated both without
weighting and with weighting by 1/d$_i$ where the $x_i$, $y_i$, and $z_i$
are the projections of d$_i$, the distance from the Galactic center
(R$_0=8$ kpc assumed). Both the weighted and unweighted solutions are
given since the true distribution probably lies somewhere in between. 
The eigenvalues of the resulting matrix are denoted e$_a$, e$_b$, and
e$_c$ and have units of kpc. 

\begin{center}
Cluster Sample with [Fe/H]$\leq-1.0$\ (weighted)\ (n=94)
\end{center}
\vspace{-0.3 in}
\begin{eqnarray}
e_{a} = 5.30^{+1.77}_{-1.11} & l = 246^{+33.6}_{-11.1} & b = 
-40.0^{+24.0}_{-18.2} \nonumber \\
e_{b} = 5.04^{+1.31}_{-0.742} & l = 44.1^{+19.2}_{-12.1} & b = 
48.0^{+12.4}_{-28.3} \\
e_{c} = 3.22^{+0.241}_{-0.464} & l = 326^{+13.9}_{-12.2} & b = 
-10.8^{+7.22}_{-13.5} \nonumber \\
& c/a = 0.61^{+0.13}_{-0.14} \nonumber 
\end{eqnarray}

\begin{center}
Cluster Sample with [Fe/H]$\leq-1.0$\ (unweighted)\ (n=94)
\end{center}
\vspace{-0.3 in}
\begin{eqnarray}
e_{a} = 19.9^{+4.03}_{-5.32} & l = 236^{+8.20}_{-24.8} & b = 
-63.2^{+25.6}_{-11.6} \nonumber \\
e_{b} = 15.5^{+2.90}_{-5.25} & l = 52.3^{+21.3}_{-8.04} & b = 
-26.8^{+15.8}_{-22.6} \\
e_{c} = 8.18^{+0.306}_{-2.01} & l = 323^{+11.7}_{-9.19} & b = 
1.53^{+4.06}_{-18.6} \nonumber \\
& c/a = 0.41^{+0.093}_{-0.091} \nonumber  
\end{eqnarray}

\noindent
As a check on the robustness of the result, solutions were also made for the 
60 
remaining clusters with [Fe/H]$\leq-1.5$.

\begin{center}
Cluster Sample with [Fe/H]$\leq-1.5$\ (weighted)\ (n=60)
\end{center}
\vspace{-0.3 in}
\begin{eqnarray}
e_{a} = 5.76^{+1.66}_{-0.721} & l = 115^{+66.3}_{-36.5} & b = 
-61.3^{+24.2}_{-4.81} \nonumber \\
e_{b} = 5.34^{+1.57}_{-1.07} & l = 60.2^{+22.9}_{-20.0} & b = 
17.3^{+29.0}_{-23.2} \\
e_{c} = 3.67^{+0.249}_{-0.511} & l = 337^{+10.6}_{-39.4} & b = 
-22.1^{+12.7}_{-10.3} \nonumber \\
& c/a = 0.64^{+0.061}_{-0.15} \nonumber  
\end{eqnarray}

\begin{center}
Cluster Sample with [Fe/H]$\leq-1.5$\ (unweighted)\ (n=60)
\end{center}
\vspace{-0.30 in}
\begin{eqnarray}
e_{a} = 19.4^{+4.67}_{-7.93} & l = 238^{+10.0}_{-33.8} & b = 
-48.5^{+10.8}_{-2.93} \nonumber \\
e_{b} = 15.1^{+4.12}_{-5.89} & l = 66.2^{+31.9}_{-8.99} & b = 
-41.3^{+11.8}_{-3.54} \\
e_{c} = 8.29^{+0.112}_{-2.11} & l = 333^{+18.2}_{-27.5} & b = 
-3.90^{+8.86}_{-23.0} \nonumber \\
& c/a = 0.43^{+0.13}_{-0.11} \nonumber  
\end{eqnarray}

\noindent
We note that for both samples the short axis of the (nearly oblate)
cluster distribution is very similar to the $\it{long}$ axis of the
kinematic solution.  In order to insure that our result is not affected by
incompleteness or uncertain absorption corrections,  solutions were also 
performed for only those
clusters whose absolute Galactic latitudes were larger than 10$^{\circ}$. 

\begin{center}
Cluster Sample with [Fe/H]$\leq-1.0$\ and $\left|b\right|>10$\ (weighted)\ 
(n=66)
\end{center}
\vspace{-0.3 in}
\begin{eqnarray}
e_{a} = 7.93^{+2.32}_{-0.618} & l = 192^{+54.2}_{-76.1} & b = 
-76.4^{+33.2}_{+0.863} \nonumber \\
e_{b} = 7.11^{+0.568}_{-1.55} & l = 58.6^{+21.9}_{-11.9} & b = 
-9.44^{+29.4}_{-33.0} \\
e_{c} = 4.64^{+0.268}_{-0.591} & l = 327^{+14.4}_{-11.3} & b = 
-9.65^{+6.99}_{-12.6} \nonumber \\
& c/a = 0.58^{+0.011}_{-0.13} \nonumber  
\end{eqnarray}

\begin{center}
Cluster Sample with [Fe/H]$\leq-1.0$\ and $\left|b\right|>10$\ (unweighted)\ 
(n=66)
\end{center}
\vspace{-0.30 in}
\begin{eqnarray}
e_{a} = 23.6^{+6.37}_{-6.36} & l = 236^{+34.1}_{-20.1} & b = 
-65.1^{+22.1}_{-7.44} \nonumber \\
e_{b} = 17.8^{+2.33}_{-4.45} & l = 50.3^{+15.3}_{-10.7} & b = 
-24.8^{+21.1}_{-20.0} \\
e_{c} = 9.43^{+0.372}_{-1.97} & l = 321^{+9.03}_{-9.78} & b = 
2.05^{+6.38}_{-11.1} \nonumber \\
& c/a = 0.40^{+0.071}_{-0.092} \nonumber  
\end{eqnarray}

The quoted uncertainties in the above solutions were determined by the 
bootstrap method. It is instructive to consider the effects on the solution 
of errors in the individual distances to each cluster. Randomly chosen errors 
drawn from a gaussian distribution with a sigma of $10\%$ of each distance
were propagated through and the mean and standard deviation of the minor axis
of the spatial distribution from 100 trials with 66 clusters was found to be 
$e_{c} = 4.67 \pm 0.0949$ towards $l = 327 \pm 1.28$ and $b = -9.91 \pm 1.02$.
The amplitudes of the other two axes are $e_{a} = 7.98 \pm 0.146$ and $e_{b} 
= 7.11 \pm 0.148$. The good agreement with solution (7) strengthens the case 
for the generally $oblate$ spatial distribution given by solution (7).

For illustration, the distribution of clusters in this new coordinate
system (solution (7)) is shown in Fig. 1 where x$^{\prime}$ are projections of
Galactocentric distance along the major axis and z$^{\prime}$ are projections 
along the minor axis.  As summarized in Table 1, the short axes of the spatial
distributions are remarkably similar to the major axes of the kinematic 
solutions of the previous section.  Furthermore, the minor axis of the 
satellite galaxy distribution is also quite similar. 

In H2000 the spatial distribution of the 34 most metal rich globular
clusters ([Fe/H]$\geq-0.7$) was also determined.  The distribution was
found to be triaxial with the long axis directed slightly off ($\sim17^
{\circ}$) the
sun-center line and the short axis now pointing to high Galactic latitude.
We repeat the process below but with one less cluster (Terzan 7). 

\begin{center}
Cluster Sample with [Fe/H]$\geq-0.7$\ (weighted)\ (n=33)
\end{center}
\vspace{-0.3 in}
\begin{eqnarray}
e_{a} = 2.02^{+0.431}_{-0.251} & l = 338^{+6.19}_{-29.0} & b = 
-6.09^{+4.63}_{-2.92} \nonumber \\
e_{b} = 1.51^{+0.273}_{-0.228} & l = 68.1^{+14.5}_{-12.7} & b = 
0.430^{+9.99}_{-9.13} \\
e_{c} = 0.761^{+0.0968}_{-0.174} & l = 334^{+65.2}_{-60.2} & b = 
83.9^{+0.907}_{-9.95} \nonumber \\
& c/a = 0.38^{+0.049}_{-0.094} \nonumber  
\end{eqnarray}

\begin{center}
Cluster Sample with [Fe/H]$\geq-0.7$\ (unweighted)\ (n=33)
\end{center}
\vspace{-0.30 in}
\begin{eqnarray}
e_{a} = 3.78^{+1.00}_{-1.29} & l = 335^{+4.86}_{-11.9} & b = 
-8.50^{+7.51}_{-2.76} \nonumber \\
e_{b} = 2.29^{+0.336}_{-0.472} & l = 65.7^{+6.05}_{-9.34} & b = 
-3.51^{+12.9}_{-10.1} \\
e_{c} = 1.11^{+0.0.0711}_{-0.413} & l = 358^{+37.5}_{-85.6} & b = 
-80.8^{+0.623}_{-9.86} \nonumber \\
& c/a = 0.29^{+0.088}_{-0.12} \nonumber  
\end{eqnarray}

\noindent
A triaxial distribution is still present, but now the major axis is 
$\sim26^{\circ}$ off of the sun-center line while the minor axis is directed 
to high latitude. Again it seems reasonable to suggest that we are observing a 
globular cluster counterpart to the `COBE' bar (c.f. Binney \& Merrifield, 
1998, p. 616).  Should future work confirm the remarkably 
close alignment between 
the major axis of the spatial distribution and the major kinematic 
dispersion axis (see Table 1), an intimate connection between halo (metal poor
clusters) and bulge (metal rich clusters) would be implied. One possible 
argument against such a connection
is the recent suggestion that bars may be less common at high redshift than 
at z=0 (van den Bergh et al. 2002). 
Further elucidation will come from the determination of the space motions 
for these metal rich clusters.

\section{Discussion}

Both the velocity ellipsoid and the spatial distribution of the most metal
poor Galactic globular clusters appear to be anisotropic.  It is
interesting that the major axis of the velocity ellipsoid is directed
towards low Galactic latitude, consistent with the direction of the short 
axis of the spatial distribution and not far from the pole of the outlying
satellite galaxy distribution.  A quantitative comparison for the
different samples is given in Table 1 where for reference we take the
major axis of the kinematic solution (1). 

\begin{deluxetable}{ccccccc}
\footnotesize
\tablenum{1}
\tablecolumns{7}
\tablecaption{Comparison of the Axes of the Various Samples}
\tablehead{
\colhead{Sample} & 
\colhead{Number} &
\colhead{Solution} &
\colhead{ Axis } & 
\multicolumn{2}{c}{Galactic Coordinates} &
\colhead{$\Delta\theta$\tablenotemark{a}}      
\\
\colhead{} & 
\colhead{n} & 
\colhead{number} &
\colhead{Major/Minor} & 
\colhead{~~~~~$l$} & 
\colhead{~~~~~$b$} &
\colhead{}
}
\startdata
Velocity ellipsoid & 38 & 1 & Major 
&~$338^{+10.5}_{-6.65}$&~$-5.65^{+13.6}_{-15.0}$& $0^{\circ}$ \nl
[Fe/H]$\leq-1.0$ & & & & & & \nl
\\ Velocity ellipsoid & 26 & 2 & Major 
&~$330^{+11.4}_{-8.11}$&~$-6.31^{+14.9}_{-14.5}$&$8.0^{\circ}\pm13$ \nl
[Fe/H]$\leq-1.5$ & & & & & & \nl
\\ Clusters & 94 & 3 & Minor 
&~$326^{+13.9}_{-12.2}$&~$-10.8^{+7.22}_{-13.5}$&$13^{\circ}\pm16$ \nl
[Fe/H]$\leq-1.0$ & & & & & & \nl
\\ Clusters & 60 & 5 & Minor 
&~$337^{+10.6}_{-39.4}$&~$-22.1^{+12.7}_{-10.3}$&$16^{\circ}\pm18$ \nl
[Fe/H]$\leq-1.5$ & & & & & & \nl
\\ Clusters & 66 & 7 & Minor 
&~$327^{+14.4}_{-11.3}$&~$-9.65^{+6.99}_{-12.6}$&$11^{\circ}\pm16$ \nl
[Fe/H]$\leq-1.0$, $|b|>10^{\circ}$ &   & &  & & & \nl
\\ Outlying Satellites & 10 & H2000 & Minor 
&~$336^{+8.28}_{-3.79}$&~$11.1^{+3.89}_{-5.05}$&$17^{\circ}\pm15$ \nl
\\ Clusters & 33 & 9 & Major 
&~$338^{+6.19}_{-29.0}$&~$-6.09^{+4.63}_{-2.92}$&$0.44^{\circ}\pm15$ \nl
[Fe/H]$\geq-0.7$ & & & & & & \nl
\enddata

\tablenotetext{a}{$\Delta\theta$ is the angular difference in degrees 
between axes of the first sample and each of the subsequent samples. }
\end{deluxetable}

There would appear to be no straight forward explanation for the
observations above. If the gravitational potential was spherical, one can 
argue that the direction of highest velocity dispersion would also be that of 
the longest spatial axis and vice versa. The above result then apparently 
rules out a spherical potential but would be consistent with one that was 
flattened in the direction of highest velocity dispersion. Recalling that 
the age-metallicity relation for the
most metal poor globular clusters is extremely steep (vandenBerg 2000), a
naive picture is that an initially flattened filament underwent a
rapid collapse tranverse to the long axis during which time satellite
galaxies, the metal poor clusters, and the  
triaxial structure discussed in H2000 were formed. Simultaneously the
Galaxy was being assembled by a slower collapse along the filament during
which time most of its angular momentum was being induced and/or 
redistributed. While triaxial 
structures are formed in cold dark matter simulations, the scenario outlined 
above may be more consistent with a warm dark matter theory for
structure formation where the lowest mass objects form from the
fragmentation of caustics (c.f. Bode, Ostriker, \& Turok 2001). 

\acknowledgments 

The author wishes to thank Dr.\ Dana Dinescu for providing the latest 
compilation of cluster space motions in convenient form, and Ray Carlberg,
Scott Tremaine, and Sidney van den Bergh for their helpful comments on an 
earlier draft. He also wishes to acknowledge financial support from an NSERC 
of Canada operating grant.

\clearpage

\begin{figure}
\plotone{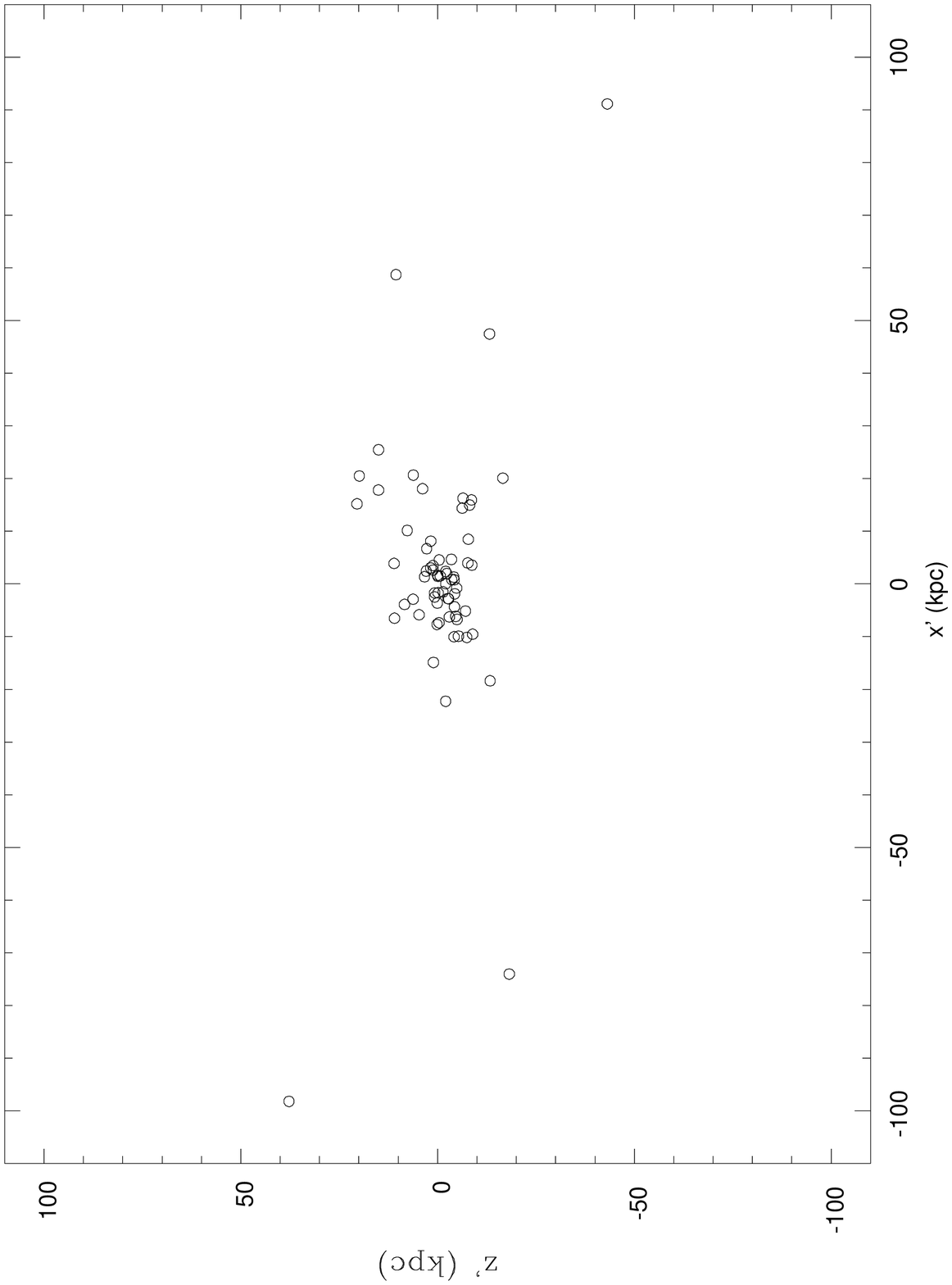}
\figcaption{
The spatial distribution of 66 globular clusters with [Fe/H]$\leq-1.0$
and $|b|>10^{\circ}$ in coordinates transformed to the major (x$^{\prime}$) and
minor (z$^{\prime}$) axes of solution (7) illustrating the flattening of the
distribution. For reference the minor axis is directed to $l=327^{\circ}$,
$b=-9.65^{\circ}$.}
\end{figure}
  
\end{document}